\documentclass[runningheads]{llncs}
\usepackage[T1]{fontenc}
\usepackage{graphicx}
\usepackage{tabularray}
\usepackage{hyperref}
\usepackage{comment}
\usepackage{subcaption}
\usepackage{makecell}
\usepackage{verbatim}
\usepackage{seqsplit}
\usepackage{glossaries}
\usepackage[capitalise,noabbrev]{cleveref}
\usepackage[table,xcdraw]{xcolor}
\usepackage{multirow}
\usepackage{array}
\usepackage{booktabs}
\usepackage{tabularx}
\UseTblrLibrary{booktabs}
\usepackage{amssymb}
\usepackage{threeparttable}
\newcolumntype{Y}{>{\raggedright\arraybackslash}X}

\begin{document}
\title{DICOMHawk: A Cyber Deception Framework for Medical Imaging Infrastructure}
\titlerunning{DICOMHawk}
%
\author{Karina Elzer\inst{1}, Alberto Mongardini\inst{1}, Ricardo Yaben\inst{1}, Georgios Theodoridis\inst{2}, \\
Alexandra Babanuta\inst{1}, Nawras Mouala\inst{1} and Emmanouil Vasilomanolakis\inst{1}}
\authorrunning{Elzer et al.}
%
\institute{Technical University of Denmark \email{\{kaelz,among,rmyl,emmva\}@dtu.dk, alexandrababanuta@gmail.com, nawras.moula3@gmail.com} \and
CSIS
\email{giorgostheo21@gmail.com}\\
}


%
\maketitle              
\begin{abstract}
\begin{sloppypar}
Cyber-attacks against exposed healthcare infrastructure threaten sensitive patient data and clinical operations, yet existing defensive tools for DICOM-based medical imaging systems provide limited interaction and are easily fingerprinted.
We introduce DICOMHawk, a cyber-deception framework that emulates DICOM and PACS services using realistic interactions, dynamically populated medical records, and embedded honeytokens. In an 86-day comparison and a 424-day deployment across multiple networks, DICOMHawk attracted more valid sessions than Dicompot, avoided honeypot detection, and captured 67 medical-related attacks from 15 unique IPs. The results show that realistic, long-term, multi-location deception improves visibility into threats targeting medical imaging systems.
\end{sloppypar}
\end{abstract}
\section{Introduction}
\label{sec:introduction}

Cyber-attacks against exposed critical systems continue to increase, causing severe disruptions and posing direct threats to public safety~\cite{COVENTRY201848}. Healthcare systems are no exception.
In Europe, health-sector incidents already exceed 300~\cite{eu_cyber_report}, while the US reports more than 710 large data breaches in 2025 alone~\cite{hipaa2026}.
Most of the reported incidents stem from vulnerabilities in exposed web applications and poor security hygiene in remote access services~\cite{cartwright2023elephant}.
However, the attack surface of healthcare institutions is far larger, consisting of interconnected Hospital, Clinical, Laboratory, and Radiology Information Systems (HIS, CIS, LIS, RIS) to manage \glspl*{ehr}.

In this study, we focus on RIS, with particular emphasis on the \gls*{dicom} protocol and the systems used to archive medical images: \gls*{dicom} viewers and workstations, \gls*{pacs}, and modality equipment.
Recently, Yazdanmehr~\cite{apilite2023} revealed the exposure of nearly 4,000 \gls*{dicom}-speaking devices accessible over the Internet without security controls and holding over 43.5 million patient records.
Our work is motivated by the discrepancy between the sensitivity of the data handled by \gls*{dicom}, the optional nature of most of its security features~\cite{dicom_sec}, and the potential reach of this issue.

While significant efforts have been made to harden the security of \gls*{dicom} and the services relying on it~\cite{eichelberg2020cybersecurity}, the work on detecting and studying attack patterns---both ongoing and new---is severely limited.
Ihanus and Kokkonen~\cite{ihanus2020modelling} propose the use of cyber deception, specifically honeypots, to enhance the cybersecurity of medical imaging systems, emphasizing their ability to detect cyber threats with a low false-positive rate.
The authors recommend deploying honeypots in healthcare networks to provide information about distinct attack phases. 
However, to the authors' knowledge, the only open source DICOM honeypot \acrfull*{dp}\footnote{\url{https://github.com/nsmfoo/dicompot}} offers limited interaction and can be detected reliably~\cite{yaben2026reallyxraymeasuringinternetexposed}, defeating the purpose of the honeypot.

This paper introduces \acrfull*{dh}, a deception framework to emulate \gls*{dicom} and \gls*{pacs} services.
In \acrshort*{dh}, attackers can interact with medical images injected with honeytokens using regular DIMSE operations, and upload their own images/malware.
After nearly a year of deployment, \acrshort*{dh} has not been identified as a honeypot by two of the biggest \gls*{cti} services (Shodan and Censys).
This study presents a unique perspective on ongoing threats to healthcare systems by comparing \acrshort*{dh} and \acrshort*{dp}, and by year-long observations of interactions and attacks involving \acrshort*{dh} across multiple networks.
Our contributions are as follows:

\begin{itemize}
   \item We propose \acrfull*{dh}, a deception framework to emulate \gls*{dicom} and \gls*{pacs} services.
   \acrshort*{dh} offers more interaction capabilities than \acrfull*{dp}, includes publicly available, anonymized medical images, and injects them with honeytokens to trace attackers back. We release our implementation as an open-source repository\footnote{\url{https://github.com/honeynet/DICOMHawk}}.
    \item We conduct a comparison between our framework and the current state-of-the-art \acrshort*{dp} during an 86-day period. Overall, \acrshort*{dh} attracted more sessions and DIMSE-C commands than \acrshort*{dp}.
   \item Additionally, we conduct a long-term measurement (424 days) of ongoing medical attacks in the wild using \acrshort*{dh} both in cloud and local networks. Our pseudo-anonymized dataset is openly accessible via Zenodo~\cite{our_dataset}.
\end{itemize}

\section{Background \& Related Work}

This section gives the basic background on \gls*{dicom} and PACS systems, while introducing previous literature on the intersection of cyber deception for healthcare infrastructure.

\subsection{DICOM \& PACS} 
\label{sec:background}

\gls*{dicom} is an international standard in the medical industry first published in 1993. It is used to exchange and store images and related data of a variety of medical image-acquisition devices (e.g. MRI, X-ray).~\cite{dicom}

DICOM represents real-world medical entities such as patients, studies, series, and images as digital objects that combine image pixel data with descriptive metadata. This metadata includes information about the patient, hospital, acquisition parameters, and equipment. 

A DICOM file typically contains Service–Object Pair (SOP) classes, which combine this sensitive data with the services allowed for it. 
The main services are DICOM Message Service Elements (DIMSE): \textit{C-ECHO} to test network connectivity. \textit{C-STORE} to transmit and store DICOM files. \textit{C-FIND} to query a DICOM database with specific attribute values. The service returns a list of requested attributes and their values for each match. On the other hand, \textit{C-GET} retrieves matching DICOM files directly over the same connection used for the query. Finally, \textit{C-MOVE} transfers matching DICOM files to a specific third-party destination.


An interaction with the DICOM protocol is called an association. An association starts with an A-ASSOCIATE request. After this the DIMSE-C commands can be used. The association ends either gracefully with an A-RELEASE or abnormally with an A-ABORT. 

The Picture Archiving and Communication System (PACS) is the system responsible for storing the files and providing network access and utilizing the protocol for transferring them. PACS also includes a user interface (often a website). It can include additional functionality such as image post-processing and diagnostic reports. In addition to digital imaging devices, it also connects to other medical systems, such as the hospital information system (HIS). PACS environments typically include two standards, DICOM and Health Level-7 (HL7). HL7 is a standard for textual data related to medical information.~\cite{seeram2019} 

\subsection{Cyber Deception for Healthcare}
\label{sec:related_work}

Deception technologies, such as network telescopes and honeypots, are widely used for threat monitoring~\cite{11106825}. The work of Beltrán-López et al.~\cite{11106825} presents several examples of cyber-deception as active defensive mechanisms for detecting and responding to attacks on these systems, such as honeypots which are fake systems designed to be attractive to attackers. However, healthcare-specific literature remains scarce.

In this space, \acrshort*{dp} stands out as the primary \gls*{dicom} honeypot with the richest set of features, mimicking the behavior of an exposed \gls*{dicom} workstation. To the best of the authors' knowledge, this is the only open source DICOM honeypot.
However, \acrshort*{dp} presents several limitations that challenge its main purpose.
\acrshort*{dp} only implements a subset of the available \gls*{dimse} operations and is not compliant with the \gls*{dicom} standard, as it refuses to share implementation details during the association process.
Lacking further configuration options to simulate other devices and systems, these two characteristics limit interactions with the honeypot and make it easily identifiable: attackers only need to verify the association messages for missing fields or attempt to store arbitrary binaries in the honeypot to confirm its identity~\cite{yaben2026reallyxraymeasuringinternetexposed}.

\acrshort*{dp} V2 extends \acrshort*{dp} with a \gls*{pacs} server, full \gls*{dimse} support, and a decoy authentication web page.
Their evaluation was conducted via simulation and a three-month deployment (May–July 2025), capturing 14,831 logs, but no \gls*{dimse} commands.~\cite{saputra2025}
However, this study presents several limitations: \textit{i)} source code is never shared, \textit{ii)} there is no deployment comparison with the original honeypot to evaluate the impact of their enhancements, \textit{iii)} the honeypot remains not compliant with the \gls*{dicom} standard, \textit{iv)} the website does not resemble \gls*{dicom} software, and \textit{v)} analysis on the web interface is superficial, making it challenging to understand whether attackers were targeting radiology systems in particular, or the correlation between observations is merely an artifact of larger scans.

While Brüggemann et al.~\cite{11623663} focus on active scanning, they mention an implementation of a low-interaction honeypot and a 9-month deployment leading to a total of 3,425 DICOM requests.

Bhargavi et al.~\cite{8776696} deployed a decoy database with \gls*{dicom} honeyrecords for anomaly detection, logging any manipulation. 
However, the framework was evaluated solely via cloud simulation rather than a real-world deployment.

There are also few studies on deception systems related to other healthcare and medical systems.
One study used honeypots---including \acrshort*{dp}---to monitor 14 protocols, in a university network over one month in 2022, capturing 61 sessions and unquantified C-FIND commands~\cite{10115073}. 
However, the broad protocol diversity limits their depth of analysis and comparative insights regarding specialized medical threats.
Moreover, Wang et al.~\cite{WANG2022108212} created and deployed 462 lightweight Internet of Medical Things (IoMT) honeypots across 22 countries to collect large-scale botnet data, primarily focusing on a novel attack-pattern categorization method.
Lastly, Shah et al.~\cite{shah2025} propose a deception network with 30 web applications used in healthcare systems that include communication over HL7 $v2$. HL7 is a protocol standard for exchange of administrative, logistical, financial health-related data.
The authors leverage synthetic datasets and canary tokens to confuse, trap, and study attackers.

Overall, the field lacks deception alternatives for healthcare systems.
While other honeypots and deception systems exist for HL7 (e.g., Medpot\footnote{https://github.com/schmalle/medpot}), \acrshort*{dp} is the only available option for radiology, a low-interaction honeypot, easily identifiable, and lacking basic features.
This paper closes the gap by introducing \acrshort*{dh}, a complete \gls*{dicom} deception system that combines and exploits known cyber-deception methods.

\section{DICOMHawk}
\label{sec:dicomhawk}

This section introduces the architecture and deception features of \acrshort*{dh}.
We build upon previous work and expand related research with new deception features to trap and identify attackers both during and after interaction with \acrshort*{dh}. \cref{tab:sota} compares \acrshort*{dh}, the original \gls*{dp} and \gls*{dp}V2.

\begin{figure}
\centering
    \includegraphics[width=0.8\columnwidth]{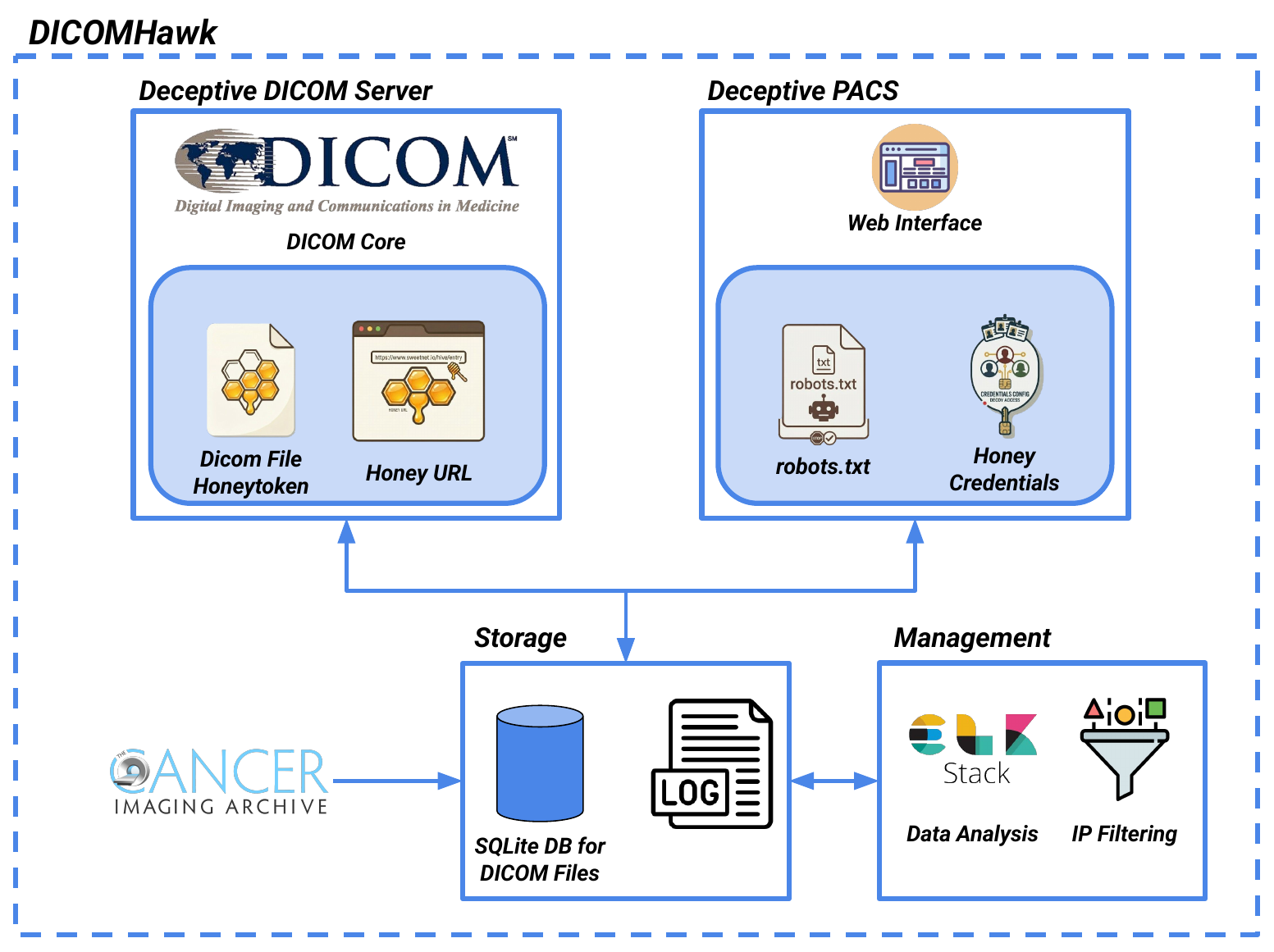}
    \caption{\acrlong*{dh}'s architecture, composed of a \gls*{dicom} server, \gls*{pacs} interface, and database storage.
    Records are dynamically collected from \gls*{tcia} and injected with PDF canary tokens and honey URLs.}
    \label{fig:architecture}
\end{figure}

\acrshort*{dh} covers a broader attack surface to study attacker behaviors by implementing the core \gls*{dimse}-C services used for verification, query, retrieval, movement, and storage, and \gls*{pacs} interfaces with planted honey credentials in commonly visited paths.
In addition, \acrshort*{dh} includes active deception mechanisms that dynamically collect public \gls*{dicom} records from \gls*{tcia}\footnote{https://www.cancerimagingarchive.net/}, populates them with fake persona information, and injects them with PDF canary tokens and honey URLs.
\cref{fig:architecture} shows the main components of the framework (i.e., \gls*{dicom} service and \gls*{pacs} interface)  and their mutual interaction with the stored honeyrecords.

\begin{table}
\centering
    \small
    \caption{Feature comparison between deception systems covered in the literature to detect and study attacks on radiology infrastructure}
    \label{tab:sota}
    
    \newcommand{\featYes}{\ensuremath{\checkmark}}
    \newcommand{\featNo}{\textendash}

\begin{threeparttable}
    \begin{tblr}{
      width=\textwidth,
      colspec={lccccccc|ccc|ccccc},
      row{even} = {bg=white}, 
      row{odd}  = {bg=gray!10},
      row{1}    = {font=\bfseries, halign=c},
      row{2}    = {cmd=\rotatebox{90}, halign=m, valign=m}, 
      column{1} = {font=\bfseries, halign=l},
      column{2-16} = {halign=c}, 
      rowsep = 1pt,          
      abovesep = 1pt,
      belowsep = 1pt,
    }
    \toprule
    \SetCell[r=2]{l} Feature & \SetCell[c=7]{c, bg=gray!20, font=\bfseries} DICOM Server & & & & & & & \SetCell[c=3]{c, bg=gray!20, font=\bfseries} PACS & & & \SetCell[c=4]{c, bg=gray!20, font=\bfseries} Active Deception & & & \\
    \cmidrule(lr){2-8} \cmidrule(lr){9-11} \cmidrule(lr){12-15}
     & A-ASSOCIATE & C-ECHO & C-GET & C-FIND & C-MOVE & C-STORE & Legit. Signature & Auth. Panel & Web DICOM Viewer & Decoy Paths & Honeycredentials & Honeyrecords & HoneyURLs & PDF Canary Tokens \\
    \midrule
    \acrshort*{dp}        & \featYes & \featYes & \featYes & \featYes & \featNo  & \featNo  & \featNo  & \featNo  & \featNo  & \featNo  & \featNo  & \featNo  & \featNo  & \featNo  \\
    \acrshort*{dp} V2\tnote{1} & \featYes & \featYes & \featYes & \featYes & \featYes & \featYes & \featNo  & \featYes & \featNo  & \featNo  & \featNo  & \featNo  & \featNo  & \featNo  \\
    \acrshort*{dh}        & \featYes & \featYes & \featYes & \featYes & \featYes & \featYes & \featYes & \featYes & \featYes & \featYes & \featYes & \featYes & \featYes & \featYes \\
    \bottomrule
    \end{tblr}
    \begin{tablenotes}
      \item[1] Source code not available.
    \end{tablenotes}
\end{threeparttable}
\end{table}

\subsection{DICOM Server}
\label{sec:dicom_server}

Existing \gls*{dicom} server implementations are not sufficiently flexible to mimic software from other vendors or libraries (e.g., dcm4che, OFFIS DCMTK or Fujifilm), apply deception features (e.g., honeyrecords), or handle/accept arbitrary binaries and commands from attackers.
Without these features and behaviors, a honeypot may still resemble a basic DICOM service, but it cannot support realistic attacker interaction or collect high-fidelity threat intelligence.
While \acrshort*{dp} solved some of these limitations, the implementation offers rather limited interaction and is easily identifiable due to incomplete support for \gls*{dimse} operations and semi-honest behavior.

\paragraph{DIMSE Operations.}

\acrshort*{dh} implements the core \gls*{dimse}-C services used for verification, query, retrieval, movement, and storage.
Attackers can store records using \texttt{C-STORE}, and retrieve them using \texttt{C-FIND} and \texttt{C-GET} queries.
Attackers can also move records within the boundaries of the storage with \texttt{C-MOVE}.
\acrshort*{dh} stores all payloads using a unique identifier and without parsing their content.

\paragraph{Legitimate Signature.}

\acrshort*{dh} modifies its signature to mimic legitimate DICOM or PACS services (e.g., Fujifilm Synapse\footnote{\url{https://healthcaresolutions-us.fujifilm.com/products/enterprise-imaging/synapse-cloud-services/}}), providing legitimate identifiers during the association process with an attacker: the system responds with configurable \texttt{AETitle}s and values included in the \texttt{UserInfo} mandatory field, modifying the \texttt{ImplementationClassUID} and \texttt{ImplementationVersionName} to masquerade the underlying library---\acrshort*{dp} does not modify or provide \texttt{UserInfo} values.
Lastly, \acrshort*{dh} logs but ignores optional fields and values sent during the association, such as \texttt{UserInfo} authentication.
\acrshort*{dh} does not currently support TLS, \texttt{AETitle} validation, or \gls*{dimse} restrictions.

\subsection{PACS Service}
\label{sec:pacs_iface}

Saputra et al. ~\cite{saputra2025} showed that most attackers targeting \gls*{dicom} services also target web interfaces where \gls*{pacs} systems, which can host web \gls*{dicom} viewers and other panels to access clinical data (e.g., for patients and healthcare personnel to access medical records, and system administrators to maintain services, users, and access).
However, the authors mention that their web interface is merely a decoy. As such, DICOMHawk incorporates a deceptive PACS web interface with authentication, \gls*{dicom} viewer and additional deceptive elements.

\paragraph{Authentication Panel.}

The authentication panel of the web interface intentionally leaks honey-credentials throughout the HTML source, actively encouraging attackers to log in and interact with the system.
When honey-credentials are used, attackers are redirected to a replacement page displaying an ``Under development'' notice, confirming unauthorized access without revealing the deception.
In addition, this panel includes validation features, allowing us to accept or reject credential dictionaries.
Brute-force attempts using random credentials are a known signature that attackers use to evade deception systems~\cite{fi18040190}.

\paragraph{Decoy Paths.}

The web interface includes additional decoy paths leaked through the \texttt{/robots.txt} path, advertising fake sensitive endpoints \texttt{/admin} and \texttt{/secure} to encourage attackers to access restricted resources. 

\paragraph{Web DICOM Viewer.}

Authenticated users can interact with a functional \gls*{dicom} viewer to browse, view, and upload \gls*{dicom} records.
The \gls*{dicom} server and \gls*{pacs} interfaces share a common database, ensuring a consistent and realistic environment across both interaction points.

\subsection{Active Deception}
\label{sec:dec_feat}


Desjardins et al.~\cite{dicomhacked} discussed how attackers can exploit \gls*{dicom} record tags to embed files with other documents and URL links as attack vectors, opening an opportunity to create honeyrecords and inject them with PDF canary tokens and honeyURLs to alert defenders when attackers access these records.
\acrshort*{dh} leverages these capabilities to serve honeyrecords through its \gls*{pacs} and \gls*{dicom} services.

\paragraph{Honeyrecords.}

\acrshort*{dh} collects records from \gls*{tcia} periodically, preventing the system from being fingerprinted through its file contents.
These records are publicly available and freely usable for educational and research purposes under the \gls*{tcia} usage policy. 
In addition, \acrshort*{dh} re-populates anonymized tags with fake persona and clinical metadata (e.g., names and hospital IDs).

\paragraph{PDF Canary Tokens.}
\begin{sloppypar}
To detect file exfiltration, we leverage the Encapsulated PDF Storage functionality of the \gls*{dicom} standard, which allows us to wrap PDF documents inside \gls*{dicom} files---this has numerous real-world applications, such as allowing radiologists to view laboratory results alongside X-rays. 
We generate PDF canary tokens by embedding invisible tracking pixels directly into \gls*{dicom} files using the \texttt{EncapsulatedDocument} tag, with the \texttt{MIMETypeOfEncapsulatedDocument} set to \texttt{application/pdf}. 
\acrshort*{dh} also overrides the \texttt{SOPClassUID} with \texttt{EncapsulatedPDFStorage} to ensure \gls*{dicom} viewers can recognize the file as an embedded PDF document and automatically render the payload. 
The canary is triggered when an attacker extracts the PDF to view its content. 
Many PDF readers automatically attempt to fetch the invisible image, sending a request to the tracking server to reveal the attacker's IP address.
\end{sloppypar}

\paragraph{HoneyURLs.}

\acrshort*{dh} also exploits the \texttt{RetrieveURL} \gls*{dicom} tag to inject a honeyURL.
This tag is normally used to provide a download link for images and remote data. 
\acrshort*{dh} injects enticing fake URLs pointing to a non-existent internal \gls*{pacs} server.
Attackers scraping metadata via \texttt{C-FIND} will encounter these URLs, and any attempt to visit them will trigger the capture of the attacker's IP address and browser metadata.
\section{Experimental Setup}
\label{sec:setup}

Our data collection spanned 424 days, from June 26, 2025, to August 23, 2026.
We initially deployed \acrshort*{dh} in two locations for testing and verification: our local institution's network and a cloud instance hosted in New York (NY).
During testing, \acrshort*{dh} used the underlying library as its signature (i.e., Pynetdicom).
\acrshort*{dh} did not log events between November and December 2025 due to networking issues.

Our testing period finalized on January 26, 2026, with the deployment of \acrshort*{dh} in an additional cloud instance in Frankfurt (FRA) to study differences in attack behaviors between European and US networks.
In addition, we deployed \acrshort*{dp} in Frankfurt cloud and local instances to compare observations between deception systems.
Therefore, we select the period between February 9 and May 5, 2026, as our comparison window (86 days) to avoid inconsistencies.

Across the entire deployment, all honeypots exposed ports 104 and 11112, with the \gls*{pacs} endpoint hosted on port 3000 (\acrshort*{dh} only). Geolocation and ISP classification for attacking IP addresses were retrieved via IPinfo\footnote{\url{https://ipinfo.io/}}.
\section{Comparison}
\label{sec:comparison}

\begin{figure}[ht]
    \centering
    \begin{subfigure}{\columnwidth}
        \centering
        \includegraphics[width=\columnwidth]{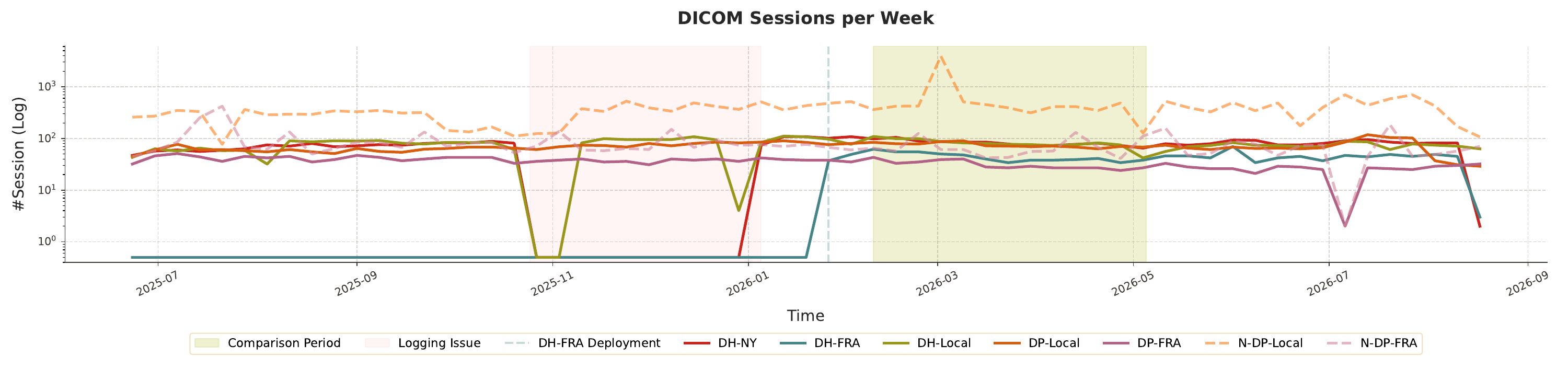}
        \caption{Sessions per week for \gls*{dicom} interactions of \acrfull*{dh} and \acrfull*{dp} with (N-DP) and without noise (DP).}
        \label{fig:dicomtimeline}
    \end{subfigure}
    
    \begin{subfigure}{\columnwidth}
        \centering
        \includegraphics[width=\columnwidth]{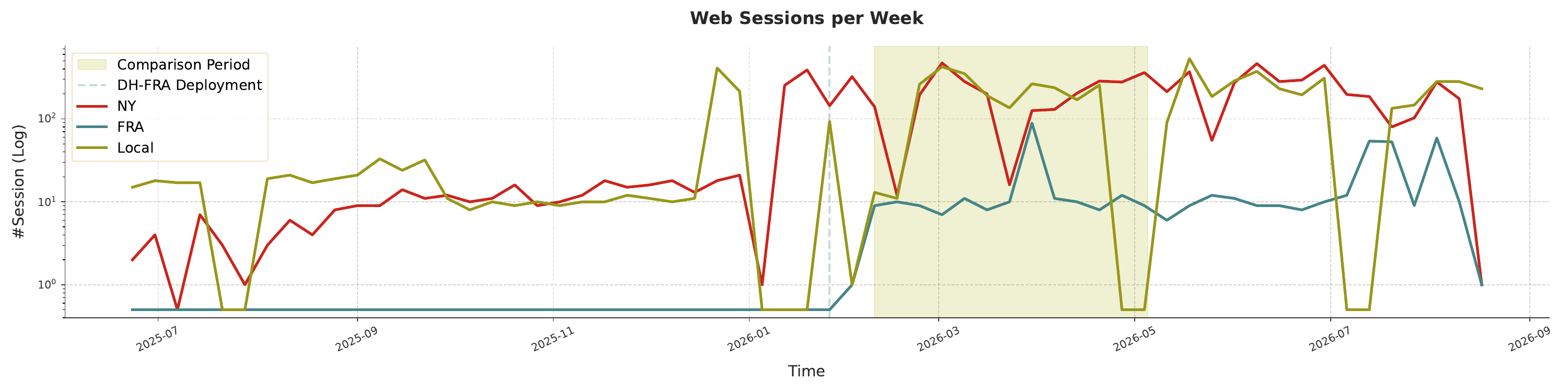}
        \caption{Sessions per week for the \gls*{pacs} Web interface for all \acrshort*{dh} instances.}
        \label{fig:webtimeline}
    \end{subfigure}
    \caption{Timelines showing the number of sessions observed per week over the year-long period by the different \gls*{dicom} honeypots and locations split by endpoints. Deployment day of the DH Frankfurt (DH-FRA) instance is marked, as well as the 86-day comparison phase (yellow) and the period impacted by the network issue (red).}
    \label{fig:timelines}
\end{figure}

This section assesses three dimensions: implementation differences (\acrshort*{dh} vs. \acrshort*{dp}), deployment locations (cloud vs. local), and interaction dynamics of \gls*{dicom} and \gls*{pacs} endpoints.
\cref{fig:dicomtimeline} illustrates the timeline categorized by honeypot endpoint and location.

\subsection{False Positives}
\label{sec:noise}

A key validation metric for any protocol-specific honeypot is its ability to filter network noise from meaningful application-level interactions. 
During this deployment, a structural divergence in logging logic between the two designs became apparent.
As shown in \cref{fig:dicomtimeline}, \acrshort*{dp} deployments (N-DP-Local and N-DP-FRA) initially appeared to log significantly more weekly sessions than \acrshort*{dh}.
However, regular traffic on the local \acrshort*{dp} instance indicated automated activity, which subsequent investigation traced to a non-\gls*{dicom} network scanner.

Conversely, \acrshort*{dh} mitigates such false positives by only logging a session after receiving and successfully parsing a valid \gls*{dicom} request. 
While this strict validation lowers raw session counts, it ensures high-confidence protocol interactions. 
To standardize the comparison, we applied a heuristic filter to the \acrshort*{dp} dataset, removing sessions lacking a client \texttt{Implementation Version Name}~\cite{yaben2026reallyxraymeasuringinternetexposed}. 
This heuristic serves as an approximation, as we observed this parameter to be absent in the generic TCP noise of the non-\gls*{dicom} network scanner; it effectively isolates valid traffic for the subsequent analysis.

\subsection{Interaction}

\cref{fig:dicomtimeline} demonstrates that \acrshort*{dh} consistently attracts more genuine \gls*{dicom} sessions than \acrshort*{dp} across all vantage points. 
As detailed in \cref{tab:hp_percentagchange}, \acrshort*{dh} recorded an average of 47.52\% more daily sessions in the cloud and 18.35\% more locally than \acrshort*{dp} (with a daily session count of 1 to 19).
This disparity suggests that higher interaction fidelity and increased fingerprinting evasion techniques significantly increase the likelihood of attacker engagement. 
Furthermore, the persistent performance gap between local and cloud deployments across both platforms indicates that vantage point and network context influence attacker behavior independently of honeypot architecture.

\begin{table}
    \small
    \centering
    \caption{Percentage change in daily sessions ($\Delta \% = \frac{D_2 - D_1}{|D_1|}$) during the 86-day period. 
    Positive values indicate an increase in $D_2$ relative to $D_1$. 
    Comparisons are grouped by Location (Cloud and Local) and Design (\acrshort*{dh} and \acrshort*{dp} w/o noise).}
    \label{tab:hp_percentagchange}
    \begin{tblr}
    {
      width=\columnwidth,
      colspec={llrr},
      row{even}  = {bg=gray!10},
      row{odd} = {bg=white},
      row{1} = {font=\bfseries, halign=l},
      column{1} = {halign=l},
      column{3,4} = {halign=r},
      rowsep = 1pt,          
      abovesep = 1pt,
      belowsep = 1pt,
    }
        \toprule
        Factor & $D_1 \rightarrow D_2$ & Mean & Median \\
        \midrule
        \SetCell[r=2]{l} Location & $Cloud_{DH}$ $\rightarrow$ $Local_{DH}$ & 107.46\% & 87.50\% \\
        & $Cloud_{DP}$ $\rightarrow$ $Local_{DP}$ & 166.59\% & 133.33\% \\
        \midrule
        \SetCell[r=2]{l} Design & $DP_{Cloud}$ $\rightarrow$ $DH_{Cloud}$ & 47.52\% & 40.00\% \\
        & $DP_{Local}$ $\rightarrow$ $ DH_{Local}$ & 18.35\% & 9.55\% \\
        \bottomrule
    \end{tblr}
\end{table}

For the comparison dataset, \acrshort*{dp} logged 4 \texttt{C-ECHO} and 4 \texttt{C-FIND} commands, while \acrshort*{dh} logged 4 \texttt{C-ECHO} and 5 \texttt{C-FIND} commands. 
As shown in \cref{tab:dicom_attacks}, \acrshort*{dp} logged 15 attacks from 3 unique IPs over its one-year deployment, compared to 27 DIMSE-C commands logged by \acrshort*{dh}. 
Two of these three IPs were also captured by \acrshort*{dh}. 
The third IP targeted the local \acrshort*{dp} instance only once during the testing period when the local \acrshort*{dh} experienced logging downtime. 
Conversely, four other IPs targeted \acrshort*{dh} but bypassed \acrshort*{dp} entirely, despite \acrshort*{dp} being active and in a comparable location.

\begin{table}
    \scriptsize
    \centering
    \caption{Attacker, target (HP+Location), query and count of observed attacks targeting the \gls*{dicom} endpoint over the year-long period.}
    \label{tab:dicom_attacks}
    \begin{tblr}{
      width=\columnwidth,
      colspec={lllXr},
      row{even}  = {bg=gray!10},
      row{odd} = {bg=white},
      row{1} = {font=\bfseries, halign=l},
      column{1} = {halign=l},
      rowsep = 1pt,          
      abovesep = 1pt,
      belowsep = 1pt,
    }
    \toprule
    HP & Attacker & Location & Query & \# \\
    \midrule
    \SetCell[r=7]{} DH & University & All & PatientName:* & 15 \\ 
    & Hosting TR & All & PatientName:* & 4 \\
    & University2 & Local & ALL STUDY & 2 \\
    & Proton & NY & PatientName: ZXZX\_IMPOSSIBLE\_NAME\_ZXZX & 2 \\
    & ISP DE & NY & ALL STUDY & 1\\
    & Comcast & Local & ALL STUDY & 2 \\
    & Comcast & Local & StudyInstanceUID:[...] & 1 \\ 
    \midrule
    \SetCell[r=3]{} DP & University & FRA+Local & PatientName:* & 12 \\
    & ISP DE & FRA & - & 2 \\
    & Hosting US & Local & PatientName: * & 1 \\
    \bottomrule
    \end{tblr}
\end{table}
  
\cref{tab:dicom_attacks} highlights a further limitation in \acrshort*{dp}'s logging fidelity: query parameters for one attack were omitted, though logs indicate that all 50 available results were retrieved. Because the source IP, \gls*{dicom} viewer, and timestamp align perfectly with an attack captured by \acrshort*{dh}, we infer this was a wildcard query for all studies. Without \acrshort*{dh}'s higher-fidelity logs, this context would have been lost. 

\subsection{Endpoints and Location}

\cref{tab:ip_overlap} shows an IP address overlap of only 11.51\% between \acrshort*{dh}'s endpoints. 
Out of 17,563 total sessions, 55.56\% targeted the \gls*{pacs} interface and 44.44\% targeted the \gls*{dicom} server. 
Accounting for data loss that primarily affected the \gls*{dicom} server, overall engagement across both protocols was roughly equal. 
Cross-location IP analysis (\cref{tab:ip_overlap}) shows minimal IP overlap, under 10\% for the \gls*{pacs} endpoint and 20\% for \gls*{dicom}. 

\begin{table}
    \small
    \centering
    \caption{Unique IP address overlap between endpoints, locations, and design of \acrfull*{dh} and \acrfull*{dp} for the 86 days.}
    \label{tab:ip_overlap}
    \begin{tblr}
    {
      width=\columnwidth,
      colspec={lrr},
      row{even}  = {bg=gray!10},
      row{odd} = {bg=white},
      row{1} = {font=\bfseries, halign=l},
      column{1} = {halign=l},
      rowsep = 1pt,          
      abovesep = 1pt,
      belowsep = 1pt,
    }
    
        \toprule
        Overlap & Count & Percentage \\
        \midrule
        \gls*{dicom} Local+NY       & 150  & 16.74\%\\
        \gls*{dicom} Local+FRA      & 98   & 10.86\%\\
        \gls*{dicom} NY+FRA         & 100  & 11.34\%\\
        \hline
        \gls*{pacs} Local+NY        & 22   & 8.46\%\\
        \gls*{pacs} Local+FRA       & 17   & 8.37\%\\
        \gls*{pacs} NY+FRA          & 18  & 7.09\%\\
        \hline
        \gls*{pacs} + \gls*{dicom}         & 158  & 11.51\%\\
        \hline
        DH + DP              & 314 & 24.86\%\\
        DH + DP Local        & 156 & 19.05\%\\
        DH + DP FRA          & 83  & 11.66\%\\
        \bottomrule
    \end{tblr}
\end{table}

Similarly, the IP distribution across deployment locations reveals minimal overlap. 
The local and New York instances exhibit the highest overlap (16.74\%). 
This lack of intersection across locations demonstrates that multi-vantage point deployments are essential for collecting a comprehensive attack dataset.
Attackers seem to rarely interact with multiple locations and endpoints, while generally preferring our local instance, as seen in \cref{tab:hp_percentagchange}.

\subsection{Scanner Evasion and Fingerprinting.}

To evaluate \acrshort*{dh}'s resistance to automated detection, we queried Shodan\footnote{\url{https://www.shodan.io/}} and Censys~\cite{10.1145/3718958.3754344}. 
Shodan tagged \acrshort*{dp} as a \texttt{"Honeypot"} but classified \acrshort*{dh} as a legitimate \texttt{"Medical"} device. 
This confirms that our architectural enhancement \gls*{dicom} services and an integrated \gls*{pacs} interface successfully evade automated fingerprinting. 
Notably, \acrshort*{dh} avoided detection even during testing while using default \texttt{pynetdicom} identifiers. 
Censys did not flag either system, indicating less aggressive \gls*{dicom} heuristics than Shodan. 
These results validate that \acrshort*{dh} achieves its primary objective: outperforming \acrshort*{dp} by remaining indistinguishable from a production deployment to both scanners and adversaries.
\section{Long-term Analysis}
\label{sec:analysis}

Given the distinct technical characteristics of the two \acrshort*{dh} endpoints, we analyze their interactions separately, with a primary focus on medical-specific attacks. 
We examine interactions with the \gls*{dicom} server first, followed by \gls*{pacs}.


\subsection{DICOM Server}

Our \gls*{dicom} server records sessions and \gls*{dimse} commands.
As summarized in \cref{tab:dicom_interactions_stacked}, the endpoint processed 9,889 sessions but only 93 \gls*{dimse} commands, indicating a heavy presence of automated scanners or reconnaissance activity. 
This includes 66 \texttt{C-ECHO} requests, which are analogous to a ping command, and we therefore categorize them as non-malicious reconnaissance.
See \cref{fig:dicomtimeline} for daily session distribution details.

\begin{table}
    \small
    \centering
    \caption{Observed \gls*{dicom} sessions and \gls*{dimse} commands from \acrshort*{dh} and unique IP counts (in parentheses) by location. 
    (A-RQ=Association Request, A-AB=Association Abort, A-RL=Association Released)}
    \label{tab:dicom_interactions_stacked}
    \begin{tblr}{
      width=\columnwidth,
      colspec={l ccc | c},
      row{even}={bg=gray!10},
      row{odd}={bg=white},
      row{1}={font=\bfseries, halign=c},
      row{10}={font=\bfseries},
      column{1}={halign=l},
      rowsep=1pt,
      abovesep=1pt,
      belowsep=1pt,
    }
        \toprule
        Event & Local & NY & FRA & Total \\
        \midrule
        A-RQ & \makecell{4,587 \\ \scriptsize (1,804 IPs)}  & \makecell{4,000 \\ \scriptsize (1,665 IPs)} & \makecell{1,302 \\ \scriptsize (965 IPs)}  & \makecell{9,889 \\ \scriptsize (2,985 IPs)} \\
        A-AB & \makecell{4,302 \\ \scriptsize (1,598 IPs)}  & \makecell{3,753 \\ \scriptsize (1,471 IPs)} & \makecell{1,161 \\ \scriptsize (841 IPs)}  & \makecell{9,216 \\ \scriptsize (2,593 IPs)} \\
        A-RL & \makecell{265 \\ \scriptsize (206 IPs)}  & \makecell{238 \\ \scriptsize (194 IPs)} & \makecell{139 \\ \scriptsize (124 IPs)}  & \makecell{642 \\ \scriptsize (393 IPs)} \\
        \midrule
        C-ECHO & \makecell{29 \\ \scriptsize (5 IPs)}  & \makecell{30 \\ \scriptsize (6 IPs)} & \makecell{7 \\ \scriptsize (5 IPs)}  & \makecell{66 \\ \scriptsize (7 IPs)} \\
        C-FIND & \makecell{11 \\ \scriptsize (4 IPs)}  & \makecell{12 \\ \scriptsize (4 IPs)} & \makecell{3 \\ \scriptsize (2 IPs)}  & \makecell{26 \\ \scriptsize (6 IPs)} \\
        C-GET & \makecell{1 \\ \scriptsize (1 IP)}  & \makecell{0 \\ \scriptsize (0 IPs)} & \makecell{0 \\ \scriptsize (0 IPs)}  & \makecell{1 \\ \scriptsize (1 IP)} \\
        C-MOVE & \makecell{0 \\ \scriptsize (0 IPs)}  & \makecell{0 \\ \scriptsize (0 IPs)} & \makecell{0 \\ \scriptsize (0 IPs)}  & \makecell{0 \\ \scriptsize (0 IPs)} \\
        C-STORE & \makecell{0 \\ \scriptsize (0 IPs)}  & \makecell{0 \\ \scriptsize (0 IPs)} & \makecell{0 \\ \scriptsize (0 IPs)}  & \makecell{0 \\ \scriptsize (0 IPs)} \\
        \midrule
        Sessions & \makecell{4,587 \\ \scriptsize (1,804 IPs)} & \makecell{4,000 \\ \scriptsize (1,665 IPs)} & \makecell{1,302 \\ \scriptsize (965 IPs)} & \makecell{9,889 \\ \scriptsize (2,985 IPs)} \\
        \bottomrule
    \end{tblr}
\end{table}

Except for \texttt{C-ECHO}, all other \gls*{dimse} commands can expose sensitive medical data; as such, we classify their execution as an attack. 
\acrshort*{dh} primarily observed \texttt{C-FIND} requests, 26 from 6 unique IPs, and a single \texttt{C-GET} request (cf. \cref{tab:dicom_interactions_stacked}). 
This prevalence is expected, as \texttt{C-FIND}'s wildcard functionality offers the fastest method to map stored patient data, while the \texttt{C-GET} requested one specific study. 
None of these attacking IPs intersected with the \gls*{pacs} endpoint.
 
Every adversary utilized \texttt{C-FIND} commands in combination with wildcard queries to request all available data at either the study or patient level (cf. \cref{tab:dicom_attacks} for attack details), with the exception of one, that requested an unusual patient name (`ZXZX\_IMPOSSIBLE\_NAME\_ZXZX'). The majority of \texttt{C-FIND} commands (57.69\%) originated from a single IP address registered to an academic institution requesting patient names. 
Given the high interaction volume, persistence across all three deployment locations, and temporal distribution (September 2025 to August 2026), this traffic likely stems from a long-term, internet-wide research study. We additionally observed a second IP connected to another academic institute.
Conversely, an attack originating from a residential ISP utilized a commercial medical viewer (\texttt{RadiAnt-2025.1}) rather than automated scripts, strongly indicating a human adversary.

\subsection{PACS Service}

The \gls*{pacs} endpoint recorded webpage visits and authentication attempts, but no successful login (cf. \cref{fig:webtimeline} for this endpoint's daily interaction). 
As detailed in \cref{tab:web_interactions_stacked}, the majority of unique IP addresses and logged events involved automated \texttt{robots.txt} requests, which are standard scanner behaviors and considered non-malicious. 
Conversely, visits to \texttt{/admin} and \texttt{/secure} pages occurred less frequently and via fewer IPs, signaling targeted reconnaissance or malicious intent. 
To isolate medical-specific threats, the subsequent analysis focuses primarily on the observed login attempts.

\begin{table}
    \small
    \centering
    \caption{Observed web events on the \acrshort*{dh}'s \gls*{pacs} endpoint and unique IP counts (in parentheses) by location.}
    \label{tab:web_interactions_stacked}
    \begin{tblr}
    {
      width=\columnwidth,
      colspec={l ccc | c},
      row{even}  = {bg=gray!10},
      row{odd} = {bg=white},
      row{1} = {font=\bfseries, halign=c},
      row{7}={font=\bfseries},
      column{1} = {halign=l},
      rowsep = 1pt,          
      abovesep = 1pt,
      belowsep = 1pt,
    }
        \toprule
        \textbf{Event} & \textbf{Local} & \textbf{NY} & \textbf{FRA} & \textbf{Total} \\
        \midrule
        /robots.txt    &  \makecell{914 \\ \scriptsize (304 IPs)}  &  \makecell{3,483 \\ \scriptsize (401 IPs)} & \makecell{1,619 \\ \scriptsize (200 IPs)}  & \makecell{6,016 \\ \scriptsize (665 IPs)} \\
        /admin         & \makecell{417 \\ \scriptsize (14 IPs)}   & \makecell{204 \\ \scriptsize (24 IPs)}  & \makecell{88 \\ \scriptsize (7 IPs)}   & \makecell{709 \\ \scriptsize (41 IPs)} \\
        /secure        & \makecell{167 \\ \scriptsize (2 IPs)}   & \makecell{19 \\ \scriptsize (3 IPs)}  & \makecell{0 \\ \scriptsize (0 IPs)}   & \makecell{186 \\ \scriptsize (4 IPs)} \\
        /admin-config  & \makecell{1 \\ \scriptsize (1 IPs)}   & \makecell{19 \\ \scriptsize (3 IPs)}  & \makecell{0 \\ \scriptsize (0 IPs)}   & \makecell{20 \\ \scriptsize (3 IPs)} \\
        Failed Login   & \makecell{12,722 \\ \scriptsize (40 IPs)} & \makecell{62,106 \\ \scriptsize (61 IPs)} & \makecell{1,786 \\ \scriptsize (7 IPs)} & \makecell{76,614 \\ \scriptsize (79 IPs)} \\
        \midrule
        Total & \makecell{14,221 \\ \scriptsize (347 IPs)} & \makecell{65,831 \\ \scriptsize (471 IPs)} & \makecell{3,493 \\ \scriptsize (210 IPs)} & \makecell{83,545 \\ \scriptsize (756 IPs)} \\
        \bottomrule
    \end{tblr}
\end{table}

\paragraph{Login Attempt Patterns.}

As \cref{tab:web_interactions_stacked} shows, failed authentication attempts were the predominant event.
Two primary patterns emerge. First, generic credentials (e.g., \textit{admin}, \textit{root}, \textit{password}) heavily dominate the dataset. Second, many passwords contain explicit contextual indicators regarding the attackers' intended targets. To categorize these attempts, we applied agglomerative clustering (\textit{distance\_threshold=0.8}) using Jaccard similarity on the password dictionary of each IP. As summarized in \cref{tab:web_targets}, this yielded 15 distinct clusters, which we mapped to broader credential themes. 

 Only 79 unique IP addresses were responsible for 76,614 login attempts, strongly indicating automated brute-force behavior. The majority of false authentication events originated from 63 IPs executing highly uniform dictionary attacks (\cref{tab:web_targets}). The observed password dictionary contained keywords such as `v2ray' and `xui' indicating attempted exploitation of 3x-ui\footnote{\url{https://github.com/mhsanaei/3x-ui}}, a web interface for the proxy project V2ray\footnote{\url{https://www.v2ray.com/en/}}. The attacks are visible in \cref{fig:webtimeline} as the recurring spikes starting at the end of 2025.

 \begin{table}
    \small
    \centering
    \caption{Categorized login attempt attacks on the \gls*{pacs} endpoint based on password dictionary clustering.}
    \label{tab:web_targets}
    \begin{tblr}
    {
      width=\columnwidth,
      colspec={lccr},
      row{even}  = {bg=gray!10},
      row{odd} = {bg=white},
      row{1} = {font=\bfseries, halign=c},
      column{1} = {halign=l},
      rowsep = 1pt,          
      abovesep = 1pt,
      belowsep = 1pt,
    }
    
        \toprule
        Target & Clusters & IPs & \# \\ 
        \midrule
        XUI/V2ray        & 1 & 63 & 76,078 \\
        CapRover         & 1 & 1 & 396\\
        Medical          & 8 & 9 & 40\\
        Other/Unidentified  & 3 & 4 & 62\\
        SQL Injection    & 1 & 1 & 32\\
        AMS30            & 1 & 1 & 6 \\
        \bottomrule
    \end{tblr}
\end{table}

Using the same methodology, we identified two additional likely targets: CapRover\footnote{\url{https://caprover.com/}} and AMS30\footnote{\url{https://www.smcworld.com/newproducts/en-id/22/ams/}}, while three other clusters used generic and non-target-specific credential dictionaries, and one cluster included SQL injection attempts.

 Interestingly, we could identify clusters related to the medical field. Even though only 0.05\% of login attempts fall into these clusters due to the intense brute-forcing attempts from the first cluster, 53\% of the unique observed brute-force attack cluster observations were medical-related. This shows our designed \gls*{pacs} endpoint successfully filters a significant amount of generic web traffic, attracting domain-specific attacks.

\paragraph{Medical-Related Login Attempts.}

Out of the 40 medical-related authentication attempts, only one targeted the local instance (Line 11, \cref{tab:medical_crendetials}) rather than the New York deployment. Notably, all nine attacking IP addresses originated from ISP address space and no commercial hosting providers; five from distinct residential ISPs in India, and another registered to the All India Institute of Medical Sciences (AIIMS). The latter IP utilized an email address containing the institute's abbreviation but registered under a public Gmail domain.

\begin{table}
    \small
    \centering
    \caption{Credential pairs of observed login attempts in medical clusters and country information of associated IP (IN=India, PK=Pakistan, GB=England, Saudi Arabia=SA, AIIMS=All India Institute of Medical Sciences).}
    \label{tab:medical_crendetials}
    \begin{tblr}
    {
      width=\columnwidth,
      colspec={lllr},
      row{even}  = {bg=gray!10},
      row{odd} = {bg=white},
      row{1} = {font=\bfseries, halign=l},
      column{1} = {halign=l},
      rowsep = 1pt,          
      abovesep = 1pt,
      belowsep = 1pt,
    }
        \toprule
        Origin & Username & Password & \# \\ 
        \midrule
        AIIMS & \nolinkurl{[redacted-email]@gmail.com} & 12345678 & 9 \\
        AIIMS & \nolinkurl{[redacted-email]@gmail.com} & lapchole & 9 \\
        ISP PK & Mahmoud & Pacs1234 & 6 \\
        ISP IN 5 & DEPT023 & pSYCH@098 & 4 \\
        ISP PK & Ellithymahmoud & Pacs1234 & 3 \\
        ISP IN  & nurse & nurse@20250 & 2 \\ 
        ISP IN 2 & Nursing & Nurse@2020 & 2 \\
        ISP IN 3 & sr\_pulmmedicine & 1234 & 2 \\
        ISP IN 4 & ortho & 1234 & 1 \\
        ISP SA & picu & 123 & 1 \\
        ISP GB & 32l4CKKtV93nvybbyeHzQ5[...] & wrong & 1 \\
        \bottomrule
    \end{tblr}
\end{table}

The low volume and distinctive nature of these attempts differentiate them from automated, dictionary-based brute-forcing. However, the underlying credentials remain rudimentary, relying on simple numeric sequences and generic medical keywords. This pattern may either reflect a low-effort manual probing attempt or highlight a systemic reliance on weak, default credentials within healthcare environments, representing a broader sector-specific security risk.
\section{Discussion}
\label{sec:discussion}

Our deception framework and deployment show clear advantages over existing solutions and research, and highlight persistent issues in health-related security.

\paragraph{DICOMHawk vs Dicompot.}

\acrshort*{dh} provides four key improvements compared to \acrshort*{dp}. First, it refines logging mechanisms to reduce false positives and ensure comprehensiveness. Second, it enhances authenticity by integrating a PACS endpoint and dynamically updating honeyrecords, preventing fingerprinting. Third, this increased realism yielded higher interaction volumes and a marginal increase in attacks. Fourth, \acrshort*{dh} injects deeper deception layers across both endpoints. Notably, embedding honeytokens within DICOM files represents a novel technique adaptable to production databases. While neither the honeytokens nor the honeyURLs were triggered, and the hidden credentials were not used, this may be related to the limited attacks observed in total. 

\paragraph{State of Attacks on Medical Systems.} 

Nevertheless, our deployment showed active engagement throughout the year, with a total of 67 medical-related attacks stemming from 15 IPs captured by DICOMHawk. While attacks focus on reconnaissance, we observe attack attempts to map stored data using a single command (\texttt{C-FIND} wildcard). Furthermore, we see interest in the Web endpoint as an easy entry point through brute-force attacks using generic credentials. 
This confirms an ongoing adversary interest in publicly exposed healthcare infrastructure. Because these specialized incidents occur far less frequently than generic internet background noise, longitudinal and multi-location measurements are essential to observe them. While the raw volume remains low, any targeted interaction with medical infrastructure carries high severity, given the extreme sensitivity of healthcare data. Continuous monitoring remains vital to identify and decode evolving medical-sector threats. 

\paragraph{Deployment Limitations.}

Although universities often host clinical systems, and modern healthcare has migrated to cloud services, such as Fujifilm Synapse Cloud, healthcare infrastructure might be more likely targeted by advanced persistent threats (APTs) pursuing highly specific targets. A deployment in dedicated healthcare environments might lead to deeper insights into the attack landscape and better protection of those critical infrastructures. 

\paragraph{Comprehensive Medical Deception Framework.}


To further enhance protection, \acrshort*{dh} iterations should aim to integrate profiles mimicking diverse DICOM/PACS servers. This includes incorporating DICOMWeb, the web-based imaging standard currently unexplored in deception research. 

With \acrshort*{dh}, we design a framework focused specifically on the DICOM protocol, analogous to Saputra et al.~\cite{saputra2025}. Following a similar idea to Shah et al.~\cite{shah2025}, future work should further investigate the possibility of a medical deception framework combining different functions, protocols, and systems. For example, integrating an advanced HL7/FHIR honeypot is a promising next step. This would replace Medpot, which remains an unmaintained, low-interaction tool restricted to legacy HL7 versions.

\section{Conclusion}
\label{sec:conclusion}

This work introduced \acrfull*{dh}, a complete \gls*{dicom} deception framework that improves on \acrfull*{dp}, the previous state-of-the-art.
The framework implements multiple deception features: a \gls*{dicom} honeypot, a \gls*{pacs} web interface, honeycredentials, honeyrecords, canary tokens, and honeyURLs.
\acrshort*{dh} collects publicly available, anonymous medical images from \gls*{tcia} and rotates them to lower identifiability.
We deployed \acrshort*{dh} and \acrshort*{dp} in multiple locations and compared interactions between the deception systems, achieving up to a 47.52\% average increase in observed daily sessions in the cloud deployment.
Our 424-day deployment of \acrshort*{dh} identified six attackers on the \gls*{dicom} honeypot, and nine in the \gls*{pacs} service.
While low in overall volume, the observed presence of medical-specific attacks presents a notable threat landscape that necessitates continuous, distributed monitoring across diverse environments.

\newglossaryentry{mri}
{
name={MRI},
description={Magnetic Resonance Imaging},
first={\glsentrydesc{mri} (\glsentrytext{mri})},
}

\newglossaryentry{ct}
{
name={CT},
description={Computed Tomography},
first={\glsentrydesc{ct} (\glsentrytext{ct})},
}

\newglossaryentry{iomt}
{
name={IoMT},
description={Internet of Medical Things},
first={\glsentrydesc{iomt} (\glsentrytext{iomt})},
}

\newglossaryentry{ce}
{
name={CE},
description={Conformité Européenne},
first={\glsentrydesc{ce} (\glsentrytext{ce})},
}

\newglossaryentry{fda}
{
name={FDA},
description={Food and Drug Administration},
first={\glsentrydesc{fda} (\glsentrytext{fda})},
}

\newglossaryentry{cert}{
name={CERT},
description={Cyber Emergency Response Team},
first={\glsentrydesc{cert} (\glsentrytext{cert})},
plural={CERTs},
descriptionplural={Cyber Emergency Response Teams},
firstplural={\glsentrydescplural{cert} (\glsentryplural{cert})},
}

\newglossaryentry{pii}
{
name={PII},
description={Personally Identifiable Information},
first={\glsentrydesc{pii} (\glsentrytext{pii})},
}

\newglossaryentry{pacs}
{
name={PACS},
description={Picture Archiving and Communication System},
first={\glsentrydesc{pacs} (\glsentrytext{pacs})},
plural={PACSs},
descriptionplural={Picture Archiving and Communication Systems},
firstplural={\glsentrydescplural{pacs} (\glsentryplural{pacs})}
}

\newglossaryentry{dicom}
{
name={DICOM},
description={Digital Imaging and Communications in Medicine},
first={ \glsentrytext{dicom} (\glsentrydesc{dicom})},
}

\newglossaryentry{ehr}
{
name={EHR},
description={Electronic Health Record},
first={\glsentrytext{ehr} (\glsentrydesc{ehr})},
plural={EHRs},
descriptionplural={Electronic Health Records},
firstplural={\glsentrydescplural{ehr} (\glsentryplural{ehr})}
}

\newglossaryentry{his}
{
name={HIS},
description={Hospital Information System},
first={\glsentrytext{his} (\glsentrydesc{his})},
}

\newglossaryentry{lis}
{
name={LIS},
description={Laboratory Information System},
first={\glsentrytext{lis} (\glsentrydesc{lis})},
}

\newglossaryentry{ris}
{
name={RIS},
description={Radiology Information System},
first={\glsentrytext{ris} (\glsentrydesc{ris})},
}

\newglossaryentry{cis}
{
name={CIS},
description={Clinical Information System},
first={\glsentrytext{cis} (\glsentrydesc{cis})},
}

\newglossaryentry{cti}
{
    name={CTI},
    description={Cyber Threat Intelligence},
    first={\glsentrydesc{cti} (\glsentrytext{cti})}
}

\newglossaryentry{tcia}
{
    name={TCIA},
    description={The Cancer Imaging Archive},
    first={\glsentrydesc{tcia} (\glsentrytext{tcia})}
}

\newglossaryentry{dimse}
{
name={DIMSE},
description={DICOM Message Service Element},
first={\glsentrydesc{dimse} (\glsentrytext{dimse})},
plural={DIMSEs},
descriptionplural={DICOM Message Service Elements},
firstplural={\glsentrydescplural{dimse} (\glsentryplural{dimse})},
}

\newacronym{dh}{DH}{DICOMHawk}
\newacronym{dp}{DP}{Dicompot}
\newacronym{sota}{SotA}{State-of-the-Art}

%
%
%
\bibliographystyle{splncs04}
\bibliography{bibliography}
\end{document}